\documentstyle [12pt] {article}
\newcommand{\nummer}[1]{\baselineskip=14pt\hskip 12 cm #1 \par}
\newcommand{\datum}[1]{\hskip 12 cm #1}
\newcommand{\titel}[1]{\Large \vskip 3 true cm\begin{center}#1\end{center}
            \normalsize\vskip 1.0 true cm}
\newcommand{\autor}[1]{\normalsize\begin{center}#1\end{center}\vskip 0.5cm}
\newcommand{\adresse}[1]{\begin{center}#1\end{center}\vskip 2 true cm}
\renewcommand{\abstract}[1]{\hfil\parbox{15.4 true cm}
{\Large ABSTRACT: \normalsize #1} \hfil \vfil \Large
\normalsize\eject}
\begin{document}
\begin{titlepage}
\nummer{\parbox[b]{4cm}{WU B 93-15\\hep-ph 9310331}}
\datum{\parbox[t]{4cm}{Sept 1993\\revised Jan 1994}}

\titel{\bf Fritzsch Texture in SUSY-SO(10) with Large Neutrino Mixing}
\autor{ YOAV ACHIMAN~\footnote{e-mail: achiman@wpts0.physik.uni-wuppertal.de}
\quad {\em and} \quad THORSTEN GREINER~\footnote{e-mail: 
greiner@wpts0.physik.uni-wuppertal.de}}
\adresse{Department of Physics \\
         University of Wuppertal \\
         Gau\ss{}str. 20, D-42097 Wuppertal \\
         Germany}
\abstract{
Fritzsch's texture is imposed on {\em all} mass matrices in a SUSY-SO(10) via 
a family $U(1)_{PQ}$ symmetry. The observed charged fermion parameters fix
the $\nu$-masses and mixing, while the later are evolved from the
GUT scale to low energies using the RG. Large $\sin^2 2{\theta}_{12}$ results.
As in a SUSY-GUT no intermediate scale is allowed, the RH-neutrino scale is the
unification one and this gives in our model $\Delta m_{12}^2 \approx 
{10}^{-10} eV^2$, in accordance with the vacuum oscillation solution to the 
solar-$\nu$ puzzle.}
\end{titlepage}
\textwidth=15.0 true cm
\textheight=22.0 true cm
\voffset -2.0cm
\hoffset -1.0cm
\parindent 0pt
\parskip 14pt
\hfuzz 1cm
\newcommand{\bye}{\end{document}}
\newcommand{\ns}{\vfill\eject}
\newcommand{\be}{\begin{equation}}
\newcommand{\ee}{\end{equation}}
\newcommand{\F}{Fritzsch\ }
\newcommand{\FT}{Fritzsch texture\ }
\newcommand{\gsim}{\lower.7ex\hbox{$\;\stackrel{\textstyle>}{\sim}\;$}}
\newcommand{\lsim}{\lower.7ex\hbox{$\;\stackrel{\textstyle<}{\sim}\;$}}
\def\NPB#1#2#3{Nucl. Phys. B {\bf#1} (19#2) #3}
\def\PLB#1#2#3{Phys. Lett. B {\bf#1} (19#2) #3}
\def\PLBold#1#2#3{Phys. Lett. {\bf#1B} (19#2) #3}
\def\PRD#1#2#3{Phys. Rev. D {\bf#1} (19#2) #3}
\def\PRL#1#2#3{Phys. Rev. Lett. {\bf#1} (19#2) #3}
\def\PRT#1#2#3{Phys. Rep. {\bf#1} C (19#2) #3}
\def\ARAA#1#2#3{Ann. Rev. Astron. Astrophys. {\bf#1} (19#2) #3}
\def\ARNP#1#2#3{Ann. Rev. Nucl. Part. Sci. {\bf#1} (19#2) #3}
\def\MODA#1#2#3{Mod. Phys. Lett. A {\bf#1} (19#2) #3}
\def\Platz{\rule[-6pt]{0pt}{24pt}}
The general interest in supersymmetric grand unified theories (SUSY-GUTs) was
revived recently in view of the observation that the gauge couplings of the
standard  model  are unified at GUT energies when the theory is supersymmetric
with $M_{SUSY} \sim M_Z - 1TeV$~\cite{sun}. At the same time the new evidence
for possible neutrino-oscillations  and the expectation that top will be
observed at FERMILAB in the near future, triggered a wave of papers
considering the supersymmetric extension of known and new fermionic mass
models~\cite{smm}~\cite{bs1}. Some of those papers~\cite{smn}~\cite{bs2} 
deal with the $\nu$-sector also. However, SUSY does not help to solve the
``Achilles heel'' of the conventional GUT see-saw models: the mass matrix of
the heavy RH neutrinos, $M_{\nu_R}$, is practically unknown in almost all
those models. The conjecture of a unite matrix, diagonal matrix or a specific
Ansatz for $M_{\nu_R}$~\cite{rev}~\cite{mm} is quite arbitrary and reduces the
reliability of such a theory. 

The new trend in model building is to reduce the number of free parameters by
using arbitrary textures. The consistent realization of such a model in terms
of SUSY-GUTs, if at all possible, requires additional complicated and/or
unnatural symmetries. One expects however, the mass matrices to be obtained
one day from the symmetry of ``the theory of everything". Hence, we think that
it will be a better strategy to look for a {\em simple} symmetry which fixes
the form of the mass matrices, even for the price of more parameters,
especially if this tells us the form of $M_{\nu_R}$ as well. 

One natural possibility to do this is to allow {\em all} mass matrices to have
the same form. If a certain texture can then account for the observed masses
and mixing it will also predict  $M_{\nu_R}$. Now, it is well known that if 
an equal form for the up and down matrices is required, only one texture can
account for the experimental data, namely the Fritzsch~\cite{f} one. The only
problem with this was resolved recently when Babu and Shafi~\cite{bs1}
showed that the Fritzsch  texture allows for a large top mass in SUSY-GUTs.
In this case, a very simple Yukawa (super-) potential suffices to give 
{\em all} mass matrices the same Fritzsch form and this can be induced via 
a special $U(1)_{PQ}$~\cite{pq} . Giving then the VEVs specific directions in
the SUSY-SO(10) space, the  neutrino sector will be completely given in terms
of known parameters of the  charged fermions.

Other special features of our model are as follows:\newline
In contrast with most neutrino mass models, the overall scale of the
RH-neutrinos, $M_R$, is not a free parameter in our case. We will take $M_R$ 
to be at $M_{GUT}$, in view of the fact that there is no room for an
intermediate scale in SUSY-GUTs. 
{\em All} mass matrices, {\em including the light neutrinos see-saw matrix},
$M_{\nu}^{light}$, are then evolved using the renormalization group (RG) 
from this scale to low energies, where they are confronted with the
experimental data.~\footnote{Babu and Shafi extended recently 
their model for leptons also~\cite{bs2}. In their paper, however, the form 
of $M_{\nu_R}$ is arbitrary and fixed by hand to have a texture different 
from the Fritzsch one.
Also, $M_R$ is a free parameter required to be in an intermediate scale and
the light neutrino masses and mixing are not evolved. Our paper is based on
the diploma thesis of one of us (T.G)~\cite{g}, presented in March 1993.}

We will give in this letter  only the essential assumptions and results, the
details will be published elsewhere~\cite{ag}.

We take the fermionic sector and the Higgs representations which contribute to
the fermionic masses to be as follows:
\begin{itemize}
\item three families of fermions in the {\bf 16} representation:
$\Psi_{\bf 16}^{(i)} \quad i=1,2,3$
\item a complex Higgs field in the representation ${\bf 10}$, 
denoted by $H_{\bf 10}$
\item three Higgs fields in the {\bf 126} representation: 
$ {\phi}_{\bf 126}^{i} \quad i=1,2,3$\ .
\end{itemize}

We need three Higgs field combinations to obtain the Fritzsch texture: 
\be
\Phi^i = H_{\bf \overline{10}}^{\phantom i} + {\phi}_{\bf \overline{126}}^i\ .
\ee
All constants in these combinations can be absorbed in the VEVs
without loss of generality. The Fritzsch form is dictated by the
following Yukawa (i.e superpotential) terms
\be
{\cal L}_Y = G_{12} \Psi_{\bf 16}^{(1)}  \Psi_{\bf 16}^{(2)} \Phi^1  +
      G_{23} \Psi_{\bf 16}^{(2)}  \Psi_{\bf 16}^{(3)} \Phi^2  +
      G_{33} \Psi_{\bf 16}^{(3)}  \Psi_{\bf 16}^{(3)} \Phi^3  + {\rm h.c} .
\ee

The Yukawa coupling constants will be made real by a redefinition of the 
relative overall phases of the fermionic representations 
$\Psi_{\bf 16}^{(i)}$\ . 
Such a potential can be obtained using a specific Peccei-Quinn like
$U(1)$~\cite{dw}~\cite{gs}.

As a result, all mass matrices will be symmetric and have the \FT, i.e. will
have the form:
\be
M_f = \left(
\begin{array}{ccc}
0 & C_f & 0 \\
C_f & 0 & B_f \\
0 & B_f & A_f
\end{array} \right)
\label{M0},
\ee
where the $A_f,\ B_f \hbox{ and } C_f \hbox { are in general complex and } f =
u, d, e, \nu, \nu_R$. They will be fixed in terms of the VEVs of the Higgs
representations.

We shall take the following VEVs directions  as denoted by their SU(5)
representation content 
\be
\begin{array}{ccccc}
< H_{\bf 10}\ > &=&  r <{\rm along\ } {\bf \bar 5}\ > &+& p <{\rm along\ } {\bf 5}\ > \\
< \Phi_{\bf 126}^1\ > &=&  t <{\rm along\ } {\bf \bar 5}\ > &+& u <{\rm along\ } {\bf 1}\ > \\
< \Phi_{\bf 126}^2\ > &=&  s <{\rm along\ } {\bf 45}\ > &+& \sigma <{\rm along\ } {\bf 1}\ > \\
< \Phi_{\bf 126}^3\ > &=&  q <{\rm along\ } {\bf \bar 5}\ > &+& w <{\rm along\ } {\bf 1}\ >.\footnotemark
\end{array}
\ee
\footnotetext{Note the similarity of our representations and directions of the VEVs to those of Harvey, Ramond and Reiss ~\cite{hrw}.
They use a different Yukawa Lagrangian, but their arguments concerning the Higgs and why the above choice of
VEVs is maintained in perturbation theory, can be applicable to our case as
well.} 

Noting now that
\be
\begin{array}{cccl}
< {\bf 5} >& \hbox{ gives a mass to }& u \hbox{ and } \nu & \hbox{ with equal weight}\ , \\
< {\bf\bar 5} >& \hbox{ gives a mass to }& d \hbox{ and } e & \hbox{ with equal weight}\ , \\
< {\bf 45} >& \hbox{ gives a mass to }& d \hbox{ and } e & \hbox{ with weights (1,-3)}\ ,
\end{array}
\ee
we can explicitly write down the entries of the mass matrices.
\renewcommand{\arraystretch}{1.5}
\be
\begin{array}{lcl}
A_d = G_{33} p &  = A_e &  \\
B_d = G_{23} (p+s) &  &  B_e = G_{23} (p-3s) \\
C_d = G_{12} p &  = C_e &  \\ 
A_u = G_{33} (r+q) &   &  A_{\nu_D} = G_{33} (r-3q) \\
B_u = G_{23} r &  = B_{\nu_D}  &      \\
C_u = G_{12} (r+t) &   & C_{\nu_D} = G_{12} (r-3t) \quad\ .
\end{array} 
\ee

Solving these equations one finds that the entries of the neutrino Dirac mass
matrix $M_{\nu_D}$, are completely given in terms of the charged fermion  mass
matrices:

\begin{eqnarray}
A_{\nu_D} & = &-3A_u + \frac{ 16A_dB_u}{3B_d + B_e} \nonumber \\
B_{\nu_D} & = &B_u \\
C_{\nu_D} & = &-3C_u +  \frac{ 16B_uC_d}{3B_d + B_e} \nonumber \quad\ .
\end{eqnarray}

The RH Majorana neutrino mass matrix, $M_{\nu_R}$,
has also the \FT\footnote{The \FT for $M_{\nu_R}$ was used already in non-SUSY
models~\cite{nuf} in a different context with different results.}
in terms of
\be
A_{\nu_R} = G_{33} w \qquad B_{\nu_R} = G_{23} \sigma \qquad
C_{\nu_R} = G_{12} u \ ,
\ee
where $w$, $\sigma$ and $u$ are the SU(5) singlet VEVs of $\phi_{\bf 126}^i $
which dictate the B-L breaking in the model. They must  be all of the same 
order of magnitude and much larger than the other VEVs. As all  $\phi_{\bf
126}^i$ come symmetrically in the Higgs potential, we shall take their VEVs
to be equal in the following, for simplicity. This cannot change the general
considerations. 

The hierarchy in entries of the RH-neutrino mass matrix comes then from the 
hierarchy of the Yukawa coupling constants
\begin{eqnarray}
\Platz
\frac{C_{\nu_R}}{B_{\nu_R}} &=&  \frac{G_{12}}{G_{23}} = 4 \frac{C_d}
	{3 B_d + B_e}\\[8pt]
\Platz
\frac{B_{\nu_R}}{A_{\nu_R}} &=& \frac{G_{23}}{G_{33}} = \frac{1}{4} 
\frac{3 B_d + B_e}{A_d} \quad . 
\end{eqnarray}

The $A_{\nu_R}$ entry dominates the RH neutrino mass matrix, $M_{\nu_R}$,
and fixes the mass scale of this matrix. This mass scale must lie in the
vicinity of $M_{GUT}$ as discussed before. Taking therefore,
\begin{eqnarray}
\Platz
M_{R}=A_{\nu_R}=M_{GUT}
\Platz
\end{eqnarray}
there is no free parameter in the neutrino sector~\footnote{Note that,
as $A_{\nu_R}=G_{33}\,\omega$ and the Yukawa coupling constant of the heavy
family must be $G_{33}\simeq 1$ to get a large top quark mass, we have
practically: $M_R \simeq \omega=u=\sigma$.}.

As in the conventional SUSY-GUTs, after the GUT symmetry breaking we have 
effectively the minimal SUSY  standard model (MSSM) between $M_{GUT}$ and 
$M_{SUSY}$. In particular, two light Higgs fields survive effectively and the
ratio of their effective VEVs is called, as usual,  $\tan\beta $. 
The mass matrices of the quarks  and leptons are then given by 
\begin{equation}
	M_{u,\nu_D} = Y_{u,\nu_D} \frac{v}{\sqrt{2}} \cos \beta
\end{equation}
for up type quarks and Dirac neutrinos and
\begin{equation}
	M_{d,e} = Y_{d,e} \frac{v}{\sqrt{2}} \sin \beta
\end{equation}
for down type quarks and charged leptons. Here, $v = 246$ GeV is the VEV of
the SM-Higgs, $\tan\beta$ is a parameter and $Y_i$ the Yukawa matrices. We
shall use in the following $A_i,B_i,C_i$ as elements of the Yukawa Matrices
normalized according to eqs. (12) or (13), respectively.

Using a redefinition of the quark fields one can eliminate all but two phases
from the quark Yukawa matrices 
\begin{eqnarray}
Y_u &=& \left(\begin{array}{ccc} 
0 & C_u & 0 \\
C_u & 0 & B_u\\
0 & B_u & A_u\\
\end{array}\right)
\nonumber\\
Y_d &=& \left(\begin{array}{ccc} 
0 & C_d \exp(i\psi)& 0 \\
C_d\exp (i\psi) & 0 & B_d \exp (i\phi)\\
0 & B_d \exp (i\phi)& A_d\\
\end{array}\right)\quad.
\end{eqnarray}

However, to obtain the best value for $|V_{cb}|$, we set as
usual~\cite{bs1}~\cite{gs}, $~\phi=0$. Note that $M_d$ and $M_e$ differ in the
$B$ entry only. Taking however,  $B_d = B_e$, as the ``naive'' SO(10)
requires, does not give the right value for $m_s$. We found that 
\be
B_e = -2 B_d
\label{BeBdrel}
\ee
results in a good fit. This relation will be used in the following discussion,
with the hope to find a group theoretical explanation for it. 
\footnote{Note, that $\phi \simeq 0$ and $B_e \simeq -2B_d$ will be anyhow
obtained as a result of the fit in the quark sector. Hence, those statements
do not effect the predictions in the neutrino sector.}

The charged lepton Yukawa matrix will be, using (\ref{BeBdrel}): 
\begin{equation}
Y_l = \left(\begin{array}{ccc} 
0 & C_d & 0 \\
C_d & 0 & -2 B_d\\
0 & -2 B_d & A_d\\
\end{array}\right)\quad.
\end{equation}
To have the see-saw mechanism, we require that 
$<{\Phi^i}_{\bf 126}>_{\bf 1}=M_R \approx M_{GUT}$.
These will give the RH neutrinos a heavy Majorana mass.
\begin{eqnarray}
\lefteqn{M^{light}_{\nu} = M^T_{\nu_D} M^{-1}_{\nu_M} M_{\nu_D}} & & 
\nonumber\\
  &=& \left(\begin{array}{ccc} 
0 & (C_{\nu_D})^2 \frac{A_d}{C_d} & 0\\
(C_{\nu_D})^2 \frac{A_d}{C_d} & (B_u - C_{\nu_D}
\frac{B_d}{4C_d} )^2 & 
\begin{array}{c}(B_u C_{\nu_D}
\frac{A_d}{C_d}\\ B_u A_{\nu_D}\\
 - A_{\nu_D} C_{\nu_D}\frac{B_d}{4 C_d}) \end{array}\\
0 & \begin{array}{c}(B_u C_{\nu_D}
\frac{A_d}{C_d}\\ B_u A_{\nu_D}\\ - A_{\nu_D}
C_{\nu_D}\frac{B_d}{4 C_d})\end{array} & (A_{\nu_D})^2
\end{array}\right)
\times M_R^{-1} \frac{v^2}{2} \sin^2 \beta .
\end{eqnarray}

We are now in a position to confront this model with experiment.\\
The entries of the quark and charged lepton Yukawa matrices can be determined
by fitting to the experimental values of the quark and lepton
masses~\cite{ckm} and  mixing~\cite{GauRGE}. 
Yet, the specific form (\ref{M0}) of the Yukawa matrices holds at the
scale of unification $M_{GUT} \approx 10^{16}$ GeV. 
It is necessary, therefore, to use RG equations for the Yukawa couplings to
evolve the matrices from  $M_{GUT}$ to the experimentally accessible scale of
$M_Z = 91$~GeV. We did this using two loop RG-equations for both the Yukawa 
and the gauge \cite{GauRGE} couplings constants. For the Yukawa
couplings the semi-analytic treatment of Barger et.~al.\ \cite{SEMI} is used. 
The RG-equations are then solved numerically using a standard
integration algorithm for differential equations.\\ 
As for the see-saw neutrino matrix,  $M^{light}_\nu$, we evolved it using one
loop RG equations as in the paper of Chankowski and
P{\l}uciennik.~\cite{SSRGE}~ \footnote{See also \cite{SSError}}\\
To scale both quark and lepton masses to our preferred scale of $M_Z$ we use
three loop QCD and one loop QED RG equations.\\ In the Yukawa sector we used
the observed masses $m_b$, $m_c$, $m_u$, $m_e$,  $m_\mu$, $m_{\tau}$ and
$V_{us}$ to fix the parameters $A_{u}$, $B_u$, $C_u$,  $A_d$, $B_e$, $C_d$ and
$\phi$.\\ 
The experimental data used in the numerical analysis are collected in Table~%
\ref{Data}. 

\begin{table}
\begin{center}
\begin{tabular}{|r@{$\,=\,$}l|}
\hline
\multicolumn{2}{|c|}{Gauge couplings \cite{gauge1} \cite{gauge2}}\\ \hline
$\alpha_1(M_Z)$ & $0.01698 \pm 0.00009$\\
$\alpha_2(M_Z)$ & $0.03364 \pm 0.0002$\\
$\alpha_3(M_Z)$ & $0.117 \pm 0.004$\\
\hline
\multicolumn{2}{|c|}{Quark masses \cite{gl}}\\ \hline
$m_u(1 \mbox{GeV})$ & $5.1 \pm 1.5 \mbox{MeV}$\\
$m_d(1 \mbox{GeV})$ & $8.9 \pm 2.6 \mbox{MeV}$\\
$m_s(1 \mbox{GeV})$ & $175 \pm 55 \mbox{GMV}$\\
$m_c(m_c)$ & $1.27 \pm 0.05 \mbox{GeV}$\\
$m_b(m_b)$ & $4.25 \pm 0.10 \mbox{GeV}$\\
\hline
\multicolumn{2}{|c|}{Lepton masses \cite{gauge1}}\\ \hline
$m_e(1 \mbox{GeV})$ & $0.4960 \mbox{MeV}$\\
$m_\mu(1 \mbox{GeV})$ & $104.57 \mbox{MeV}$\\
$m_\tau(1 \mbox{GeV})$ & $1.7835 \mbox{GeV}$\\
\hline
\multicolumn{2}{|c|}{CKM matrix entries \cite{ckm}}\\ \hline
$V_{us}$ & 0.218 --- 0.224\\
$V_{ub}$ & 0.001 --- 0.007\\
$V_{cb}$ & 0.030 --- 0.058\\
\hline
\end{tabular}
\end{center}
\caption{The experimental data used in the numerical analysis}
\label{Data}
\end{table}
Let us now summarize the results of the numerical analysis.\\
Keeping the SUSY breaking scale $M_{SUSY}$ fixed, we can determine the
unification scale $M_X$ and the unified coupling $\alpha(M_X)$ from the
requirement that $\alpha_1(M_X) = \alpha_2(M_X)$. For a SUSY breaking scale,
$M_{SUSY} = 200$ GeV, we found $M_X \approx 1.6\times 10^{16}$ GeV and
$\alpha(M_X) \approx 1/25$. This gives $\alpha_3(M_Z) \simeq 0.116$, in
excellent agreement with the experimental data.

Only a tiny range for  $\tan\beta$, is allowed for the  present experimental 
value of\\ 
$|V_{cb}| = 0.030$ --- $0.058$\\
\centerline{ $ 62.7 \leq \tan \beta \leq 64$ . }

The model predicts therefore the top mass as well as the mass of the 
s-quark.\\ 
Fig.~1 shows the dependence of $m_t(m_t)$ on
$|V_{cb}|$. We find the top quark mass to be in the range\\
\centerline{ $ 94 \leq m_t \leq 157$ . } 

The strange quark mass varies only weakly with $m_t(m_t)$. We find
$m_s(1\,\mbox{GeV}) = 124 \pm 5 \,\mbox{MeV}$ near to the lower bound of
the experimental data. Note that this prediction will change once 
$B_e \ne -2B_d$.

For the neutrino mixing parameters $\sin^2 2\theta_{ij}$ we find
\begin{eqnarray}
\sin^2 2\theta_{12} &=& 0.991 \\
\sin^2 2\theta_{13} &=& 2\times 10^{-3} \;\mbox{---}\; 0.01\\
\sin^2 2\theta_{23} &=& 0.32 \;\mbox{---}\;  1.0  .
\end{eqnarray}
The masses of the light neutrinos are as follows
\begin{eqnarray}
m_1 &=& 2.4\times 10^{-5}\,\mbox{eV} \;\mbox{---} 5.3\times
10^{-5}\,\mbox{eV}\\
m_2 &=& 2.6\times 10^{-5}\,\mbox{eV} \;\mbox{---} 5.8\times
10^{-5}\,\mbox{eV}\\
m_3 &=& 4.0\times 10^{-3}\,\mbox{eV} \;\mbox{---} 1.9\times
10^{-2}\,\mbox{eV},
\end{eqnarray}
resulting in the squared mass differences 
\begin{eqnarray}
	\Delta m^2_{12} &=& ( 1.0\;\mbox{---}\;5.9) \times 10^{-10}
		\mbox{eV}^2\\
	\Delta m^2_{23} &=& ( 1.6\;\mbox{---}\;37) \times 10^{-5}
		\mbox{eV}^2.
\end{eqnarray}

We see that the $\nu_e$,$\nu_{\mu}$ mixing and masses lie in the right range
to solve the solar $\nu$-problem~\cite{b}~\cite{s} via vacuum oscillations
~\cite{bp}~\cite{kp}. This solution is called sometimes``just so"~\cite{gk}.

We can therefore conclude as follows:\\
A very simple superpotential, invariant under  $U(1)_{PQ}$, gives {\em all}
mass matrices the  Fritzsch texture and hence predicts the form of
$M_{\nu_R}$.
The overall scale of $M_{\nu_R}$ is taken to be $M_{GUT}$ as no
intermediate scale is natural in SUSY-GUTs. The neutrino sector is then 
completely fixed, in our model, in terms of known parameters of the charged
fermions. All masses and mixing, including the see-saw matrix
$M_{\nu}^{light}$, are evolved using the full renormalization group equations.
As a result we obtain large $\nu$-mixing and 
$\Delta m_{12}^2 \approx 10^{-10} eV^2$,
as is required by the ``just so" solution to the solar-$\nu$ puzzle. 

We thank Daniel Wicke for discussions. One of us (T.G) would like to
acknowledge a graduate scholarship of the Deutsche Forschungsgemeinschaft.


{\Large\bf Figure Captions}
\begin{itemize}
\item[Fig. 1 :] $m_t(m_t)$ dependence on $|V_{cb}|$.
\item[Fig. 2 :] The neutrino masses as a function of $m_t(m_t)$.
\item[Fig. 3 :] The dependence of the squared mass difference $\Delta
m^2_{12}$ on the mixing parameter $\sin^2 2\theta_{12}$.
\end{itemize}

\bye